\documentclass[
aip,
 amsmath,amssymb,
 reprint,%
]{revtex4-1}

\usepackage{graphicx}
\usepackage{dcolumn}
\usepackage{bm}
\usepackage[normalem]{ulem}
\usepackage[utf8]{inputenc}
\usepackage[T1]{fontenc}
\usepackage{mathptmx}
\usepackage{etoolbox}
\usepackage{xcolor}
\usepackage{filecontents}
\usepackage{gensymb}
\usepackage{comment}

\makeatletter
\def\@email#1#2{%
 \endgroup
 \patchcmd{\titleblock@produce}
  {\frontmatter@RRAPformat}
  {\frontmatter@RRAPformat{\produce@RRAP{*#1\href{mailto:#2}{#2}}}\frontmatter@RRAPformat}
  {}{}
}%

\makeatother
\begin{document}

\preprint{AIP/123-QED}

\title{Giant Hall effect in two-dimensional CoSi$_2$ granular arrays}

\author{Elica Anne Heredia}
\affiliation{International College of Semiconductor Technology, National Yang Ming Chiao Tung University, Hsinchu 30010, Taiwan}

\author{Shao-Pin Chiu}
\affiliation{Department of Physics, Fu Jen Catholic University, Taipei 24205, Taiwan}

\author{Ba-Anh-Vu Nguyen}
\affiliation{Department of Electrophysics, National Yang Ming Chiao Tung University, Hsinchu 30010, Taiwan}

\author{Ruey-Tay Wang}
\affiliation{Department of Electrophysics, National Yang Ming Chiao Tung University, Hsinchu 30010, Taiwan}l

\author{Chih-Yuan Wu}
\affiliation{Department of Physics, Fu Jen Catholic University, Taipei 24205, Taiwan}

\author{Sheng-Shiuan Yeh}
\altaffiliation [Authors to whom correspondence should be addressed: ] {ssyeh@nycu.edu.tw (S.S.Y.) and jjlin@nycu.edu.tw (J.J.L.)}

\affiliation{International College of Semiconductor Technology, National Yang Ming Chiao Tung University, Hsinchu 30010, Taiwan}\

\affiliation{Center for Emergent Functional Matter Science, National Yang Ming Chiao Tung University, Hsinchu 30010, Taiwan}

\author{Juhn-Jong Lin}
\altaffiliation [Authors to whom correspondence should be addressed: ] {ssyeh@nycu.edu.tw (S.S.Y.) and jjlin@nycu.edu.tw (J.J.L.)}

\affiliation{Department of Electrophysics, National Yang Ming Chiao Tung University, Hsinchu 30010, Taiwan}

\date{\today}

\begin{abstract}

Granular metals offer tailorable electronic properties and play crucial roles in device and sensor applications. We have fabricated a series of nonmagnetic granular CoSi$_2$ thin films and studied the Hall effect and transport properties. We observed a two orders of magnitude enhancement in the Hall coefficient in films fall slightly above the metal-insulator transition. This giant Hall effect (GHE) is ascribed to the local quantum-interference effect induced reduction of the charge carriers. Transmission electron microscopy images and transport properties indicate that our films form two-dimensional granular arrays. The GHE may provide useful and sensitive applications.
\end{abstract}

\maketitle

\section{Introduction}
Granular metals, $M_xI_{1-x}$, are composites comprising metals ($M$) and insulators ($I$), where $x$ denotes the volume or area fraction of $M$.\cite{Abeles1975} In granular composites, the size of metallic particles typically ranges from a few to hundreds of nanometers. Granular metals are inhomogeneously disordered, which forms either three-dimensional (3D) or two-dimensional (2D) arrays, depending on the metal grain size relative to film thickness. They can be considered as artificial solids with engineerable electronic and optical characteristics.\cite{Beloborodov2007} They serve as not only useful materials for nanotechnology applications but also controllable systems for exploring the quantum-interference and electron-electron interaction (EEI) effects\cite{Beloborodov2007,Zhang2011} as well as the percolation problem.\cite{Abeles1975} In practice, research on granular composites has led to the development of chemical sensors,\cite{Lith2007,Muller2011} strain gauges,\cite{Herrmann2007,Huth2010} temperature-insensitive resistors,\cite{Yajadda2011} etc. 

The Hall coefficient $R_H$ provides important information about the charge carrier density ($n^\ast$) and polarity in a conductor. The knowledge of $R_H$ of granular metals, especially its behavior in the vicinity of the percolation threshold or the metal-insulator transition (MIT), is of particular interest. According to the classical percolation theory, $R_H$ is sensitive to the dimensionality of the system.\cite{Shklovskii1977,Bergman1983,Straley1980,Juretschke1956} In 3D, the theory predicts that as $x$ decreases from 1, $R_H$ monotonically increases and takes a maximum value at the classical percolation threshold ($x_c$). In 2D, the theory predicts that $R_H$ retains the same value of that of the pure metal in the entire metallic regime ($x \geq x_c$). These classical predictions for the 3D case\cite{Bandyopadhyay1982} and 2D case\cite{Palevski1984} were previously observed in experiments. In this work, we report a finding of an enhancement by a factor $\approx$\,100 of $R_H$ as $x$ approaches the quantum percolation threshold, denoted by $x_q$ ($> x_c$), in a series of CoSi$_2$ thin films which form 2D granular arrays. To our knowledge, such giant Hall effect (GHE) in 2D granular arrays has never been previously found. We interpret that the GHE arises from a reduction of $n^\ast$ due to the carrier localization induced by the local quantum interference effect in an inhomogeneously disordered systems containing rich microstructures.\cite{Zhang2001} We mention that a continuous CoSi$_2$ film is a good metal with its temperature ($T$) behavior of resistivity well described by the Boltzmann transport equation, until it undergoes superconducting at about 1.5 K.\cite{Chiu2024}

\section{Experimental Method}

A series of CoSi$_2$ thin films with various thicknesses, and thus grain sizes, were grown on $\approx$\,300-nm-thick SiO$_2$ capped Si substrates. A Si layer with thickness $t_{\rm Si}$ was deposited on the substrate via thermal evaporation in a high vacuum ($\sim 2$$\times$$10^{-6}$ torr), followed by the evaporation deposition of a Co layer of thickness $t_{\rm Co}$. The ratio $t_{\rm Si} = 3.6\,t_{\rm Co}$ was chosen to ensure a complete reaction between Co and Si to form the CoSi$_2$ phase in subsequent thermal annealing process.\cite{Chiu2017} After the thermal annealing process, the thickness $t$ of the resulted CoSi$_2$ film was expected to be $t \simeq$ 3.5\,$t_{\rm Co}$, as previously established.\cite{Ommen1988} This series of films, called group A, had thickness $t$ in the range $11<t<52$ nm and form a 2D granular array. A second series of CoSi$_2$ films, called group B, with $33<t<105$ nm was grown via the deposition of a $t_{\rm Co}$ thick Co layer directly on a high-purity Si(100) substrate.  The as-deposited groups A and B films were annealed in a high vacuum ($\sim 1$$\times$$10^{-6}$ torr) at 600--800$^\circ$C for 1 h to form the CoSi$_2$ phase. The CoSi$_2$ structure was polycrystalline in group A, while nominally epitaxial in group B.\cite{Chiu2017} To facilitate transport measurements, a metal shadow mask was used during the deposition process to define a Hall bar geometry of 1 mm wide and 1 mm long. The relevant parameters of our films are listed in Table \ref{table_grain}.

The topology of the CoSi$_2$ film surface was characterized by the atomic force microscopy (AFM) (DFM SPA 400). The cross-sectional transmission electron microscopy (TEM) studies were performed using a high resolution transmission electron microscope (JEOL JEM-F200). Four-probe electrical and Hall effect measurements were carried out using a closed-cycle refrigerator and a $^3$He fridge.

\section{Results and Discussion}
\subsection{Giant Hall effect in 2D CoSi$_2$ granular arrays}

\begin{figure*}
	\centering
	\includegraphics[width=1.0\linewidth]{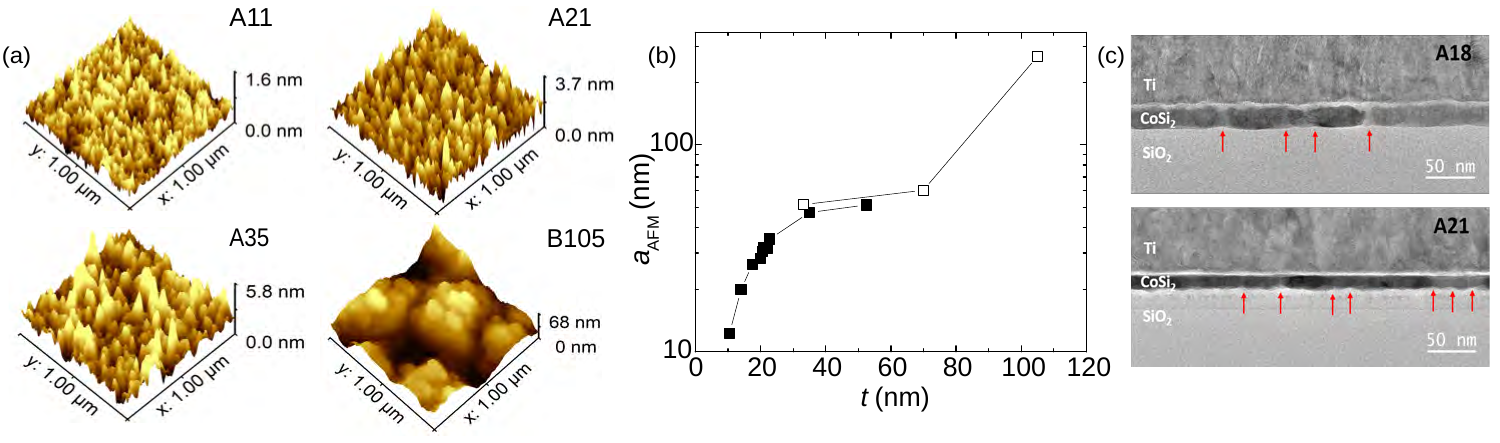} 
    \caption{(a) AFM images for four films, as indicated. (b) Average grain size $a_{\rm AFM}$ as a function of $t$. Solid (open) symbols denote group A (B) films. (c) Cross-sectional TEM images for films A18 and A21. Red arrows indicate the boundaries between neighboring CoSi$_2$ grains.}
	
	\label{fig_1}
\end{figure*}

In Fig. \ref{fig_1}(a), the surface topography for three group A and one group B CoSi$_2$ films measured by AFM reveals mountain peak-like profiles. Assuming these peaks reflect the profiles of the surfaces of the constituent grains and grains are spherical, we estimate the grain size ($a_{\rm AFM}$) in each film using the Gwyddion software (version 2.61). The data gathered from the AFM images were first leveled by mean plane subtraction, followed by row aligning using the median and matching methods. Horizontal scars or strokes were corrected to ensure accuracy, followed by color mapping to aid in selecting the appropriate range for analysis. Individual grains were then identified by selecting a threshold height of 50 percent, and all grain sizes within a 1-$\mu$m$^{2}$ area were measured and averaged. The obtained $a_{\rm AFM}$ values are listed in Table \ref{table_grain}. 

Figure \ref{fig_1}(b) shows the variation of $a_{\rm AFM}$ with $t$. For the B105 film, we obtain $a_{\rm AFM} \approx$\,330 nm, which is close to that determined from the scanning electron microscopy (SEM) and the TEM images.\cite{Chiou2015} For group A films, as $t$ increases from 11 to 52 nm, $a_{\rm AFM}$ increases from 12 to $\approx$\,50 nm. For group B films, as $t$ increases from 33 to 105 nm, $a_{\rm AFM}$ increases from $\approx$\,50 to $\approx$\,330 nm. We note that, in every film, $a_{\rm AFM} \gtrsim t$. This implies that all of our films are constituted of a single layer of CoSi$_2$ grains. The TEM images in Fig. 1(c) and those from our previous work\cite{Chiou2015} also confirm this assertion. Thus, our CoSi$_2$ films form 2D granular arrays (see further discussion below). 

\begin{figure}
	\centering
	\includegraphics[width=0.9\linewidth]{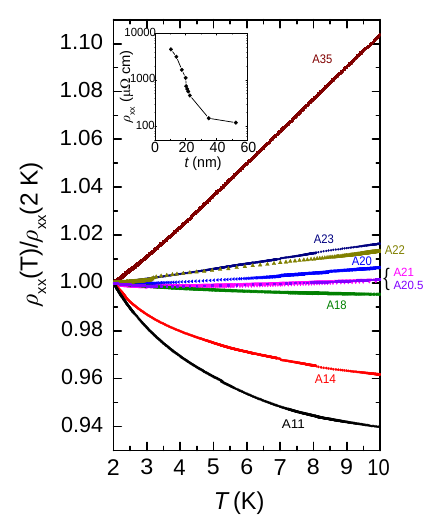}
	\caption{Normalized longitudinal resistivity $\rho_{xx} (T)/ \rho_{xx} \rm{(2\,K)}$ as a function of $T$ for group A films, as indicated. Inset: Variation of $\rho_{xx}$(2\,K) with $t$ for group A films.  
	}
	\label{fig_2}
\end{figure}

We have measured the longitudinal resistivity $\rho_{xx}$ and $R_H$ of our films and obtained $n^\ast$ from $n^\ast=1/e R_H$. The Hall effect measurements indicate that the charge carriers are holes, in consistency with our previous results\cite{Chiu2017,Chiu2024} and band structure calculations.\cite{Mattheiss1988,Newcombe1988} The elastic electron mean free path $\ell_e$ is then calculated through $\rho_{xx}^{-1} = k_F^2e^2\ell_e/(3\pi^2\hbar)$, with the Fermi wavenumber $k_F = (3 \pi^2 n^\ast)^{1/3}$. The value of the product $k_F \ell_e$ for every film is listed in Table \ref{table_grain}. Figure \ref{fig_2} shows the normalized resistivity, $\rho_{xx}(T)/\rho_{xx}(\rm 2\,K)$, as a function of $T$ for several group A films between 2 and 10 K. We see that those films having $t \gtrsim 20$ nm exhibit metallic behavior, \textit{i.e.}, $\rho_{xx}$ increases with increasing $T$; while those films having $t \lesssim 18$ nm reveal insulating behavior, \textit{i.e.}, $\rho_{xx}$ decreases with increasing $T$. Thus, the metal-insulator transition occurs around $t \simeq 19$ nm. In other words, for films with $t \gtrsim 20$ nm, the neighboring CoSi$_2$ grains are geometrically connected, forming a percolating conduction array. For films with $t \lesssim 18$ nm, the CoSi$_2$ grains are geometrically disconnected. The inset of Fig. 2 shows $\rho_{xx}$(2\,K) as a function of $t$. As $t$ decreases from 53 to 11 nm, $\rho_{xx}$(2\,K) increases rapidly from 131 to 4230 $\mu\Omega$ cm, indicating that the granularity plays an increasingly important role especially when $t$ decreases to below about 20 nm. For comparison, $\rho_{xx}$(2\,K) = 4.55, 2.70 and 2.50 $\mu\Omega$ cm for nominally continuous films B33, B70 and B105, respectively.

\begin{table}
\caption{\label{table_grain}
Relevant parameters of CoSi$_2$ films. $t$ is film thickness, $a_{\rm AFM}$ is mean diameter of CoSi$_2$ grains, $k_F$ is Fermi wavenumber, $\ell_e$ is elastic mean free path, and $L_\varphi$ is dephasing length.}

\begin{ruledtabular}
\begin{tabular}{lcccc}
Film & $t$ (nm) & $a_{\rm AFM}$ (nm) & $k_F \ell_e$ & $L_\varphi$(2\,K) (nm) \\ 
\hline 
A11   & 11   & 12  & 1.0 & 40 \\
A14   & 14   & 20  & 1.5 & 38 \\
A18   & 18   & 26  & 3.1 & 47 \\
A20   & 20   & 28  & 5.1 & 40 \\
A20.5 & 20.5 & 30  & 8.3 & -- \\
A21   & 21   & 32  & 10  & 57 \\
A22   & 22   & 32  & 10  & -- \\
A23   & 23   & 35  & 11  & 69 \\
A35   & 35   & 47  & 21  & 100 \\
A52   & 52   & 51  & 24  & 130 \\
B33   & 33   & 52  & 310 & 690 \\
B70   & 70   & 60  & 510 & 1300 \\
B105  & 105  & 330 & 540 & 1400 \\

\end{tabular}
\end{ruledtabular}
\end{table}

\begin{figure}
	\centering
\includegraphics[width=1\linewidth]{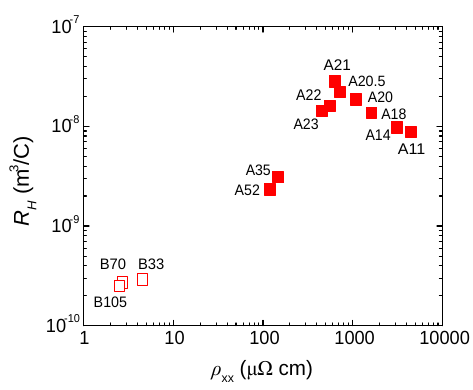}
	\caption{Hall coefficient $R_H$ as a function of $\rho_{xx}$ for granular group A and nominally epitaxial group B films at $T$ = 2 K. }
    
	\label{fig_3}
\end{figure}

Figure \ref{fig_3} shows $R_H$ as a function of  $\rho_{xx}$ at $T$ = 2 K. The thickest film B105 has the lowest Hall coefficient $R_H \simeq$ 2.5$\times$$10^{-10}$ m$^3$/C, in agreement with the previous result.\cite{Radermacher1993,VanOmmen1990} This film will serve as our reference film with the $R_H$ value for a good CoSi$_2$ metal. Note that for films A21--A52, $R_H$ monotonically increases with increasing $\rho_{xx}$, reaching a maximum value of $\approx$ 2.8$\times$$10^{-8}$ m$^3$/C (corresponding to $n^\ast \approx$ 2.2$\times$$10^{26}$ m$^{-3}$) in film A21. This is a film falling just above the MIT. Interestingly, as $\rho_{xx}$ ($t$) further increases (decreases), $R_H$ progressively decreases to $\approx$ 8.7$\times$$10^{-9}$ m$^3$/C in film A11, which falls below the MIT. We emphasize that the $R_H$ value of film A21 is $\approx$\,100 times larger than that of film B105. This is the GHE.\cite{Wan2002} Recall that according to the classical percolation theory,\cite{Zhang2001,Wu2010} $R_H(x \rightarrow x_c)$ will be enhanced by a factor $\sim (t/a)^{\tilde{g}/\nu}$, where $\tilde{g}$ is the critical exponent of Hall resistivity, $\nu$ is the exponent of the correlation length, and $a$ is the grain size. However, for a 2D percolation system, $\tilde{g}=0$ (Ref. \onlinecite{Straley1980}) and $\nu = 4/3$ (Ref. \onlinecite{Bergman1992}), thus $R_H$ should remain a constant in the entire metallic regime ($x \lesssim x_c$), as mentioned.\cite{Shklovskii1977,Bergman1983,Straley1980,Juretschke1956} This prediction was confirmed by a previous study of 2D granular Au films.\cite{Palevski1984} On the other hand, $\tilde{g}$ = 0.4 and $\nu$ = 0.9 in 3D, thus an enhancement of $R_H(x_c)$ by a factor of $\sim$\,10 is expected for, \textit{e.g.}, a thick granular film with $t$ = 1 $\mu$m and $a$ = 10 nm. The GHE means that the measured $R_H(x \rightarrow x_q)$ value is orders of magnitude larger than the classical $R_H(x \rightarrow x_c)$ value.

Wan and Sheng\cite{Wan2002} have developed a quantum percolation theory which considers the wave nature of the charge carriers in a 3D nonmagnetic granular system. At low $T$, when the electron dephasing length $L_\varphi$ is larger than the feature size ($\xi$) of the microstructures, the carrier wavefunctions will undergo multiple scattering in the random percolating clusters of the conducting channels until a phase-breaking event takes place. Such multiple coherent scattering leads to a local quantum-interference effect which causes significant localization of carrier wavefunctions as $x \rightarrow x_q$. Consequently, $n^\ast$ will be greatly reduced, which in turn gives rise to a greatly enhanced $R_H$ at $x \simeq x_q$ $(> x_c)$. This theory has successfully explained the GHE observed in 3D Cu$_x$(SiO$_2$)$_{1-x}$ composites\cite{Zhang2001} and Mo$_x$(SnO$_2$)$_{1-x}$ composites,\cite{Wu2010} where an enhancement by a factor of nearly three orders of magnitude in $R_H (x \rightarrow x_q)$ was observed. Due to this local quantum interference effect, which is absent in the classical percolation theory, $R_H$ peaks at a metal volume fraction $x_q > x_c$. The GHE in 2D has not been theoretically treated in the literature.

To clarify the underlying physics for the GHE observed in our group A films, we have measured the weak-localization induced magnetoresistance\cite{Lin2002} at low $T$ and extracted $L_\varphi$ for each film, see Table \ref{table_grain}. The $L_\varphi$(2\,K) values in group B nominally epitaxial films are much longer than those in group A polycrystalline films, as expected. In both cases, $L_\varphi$ decreases with decreasing $t$, due to an increasing dephasing rate as the disorder is increased with decreasing $t$ and decreasing $a_{\rm AFM}$. Most important, we obtain $L_\varphi$(2\,K) $> \xi \sim a_{\rm AFM} > t$ in all films. Consequently, multiple coherent electron scattering within a characteristic area of $L_\varphi^2$ leads to a quasi-2D local quantum interference effect, resulting in a reduced $n^\ast$ and an enhanced $R_H$.\cite{Zhang2001, Wan2002}

In the percolation theory, the relevant parameter is the metal volume (area) fraction $x$, which cannot be accurately determined in our films. Nevertheless, we know that $x_c$ occurs around $t \approx$ 19 nm where the MIT takes place. On the other hand, we may assume $x_q$ occur around $t \approx$ 21 nm where the $R_H$ value peaks. This is in consistency with the above quantum percolation theory prediction that $x_q > x_c$.\cite{Wan2002} At this critical thickness, one expects the film have a value of the product $k_F\ell_e \sim 2\pi$, \textit{i.e.}, the Ioffe-Regel criterion, see Table \ref{table_grain}. Thus, our observation of the 2D version of the GHE is satisfactorily understood.

\subsection{Transport in 2D CoSi$_2$ granular arrays}

\begin{figure}
	\centering
\includegraphics[width=0.9\linewidth]{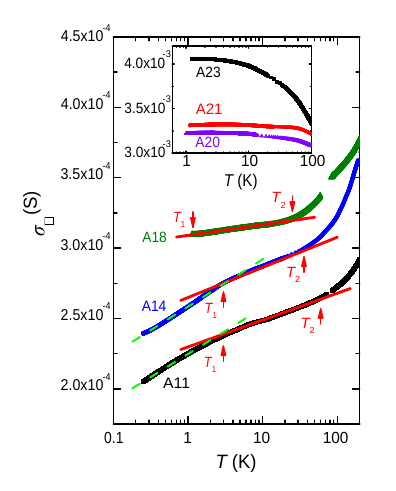} 
	\caption{Sheet conductivity $\sigma_\square$ as a function of $T$ for films A11, A14 and A18. The data of A14 (A18) are offset by $-2$$\times$$10^{-4}$ ($-7.6$$\times$$10^{-4}$) S, for clarity. Inset: $\sigma_\square$ as a function of $T$ for films A20, A21 and A23. The data of A20 (A23) are offset by $1.4$$\times$$10^{-3}$ ($-9$$\times$$10^{-4}$) S, for clarity.}
	\label{fig_4}
\end{figure}

To further clarify the occurrence of the GHE and the array dimensionality ($\tilde{d}$) of group A films, we study their longitudinal transport behavior. Figure \ref{fig_4} shows the sheet conductivity $\sigma_\square$ as a function of $T$ for films A11, A14 and A18, which have $x<x_c$. We find a $\sigma_\square \propto {\rm ln}T$ dependence in the temperature regime $T_1 < T < T_2$,  with $T_1 \approx$ 3, 3 and 1.2 K, and $T_2 \approx$ 60, 37 and 16 K, for films A11, A14 and A18, respectively. This $\ln T$ behavior does not originate from the 2D EEI effect of Altshuler and Aronov (AA),\cite{Altshuler1985} which considers homogeneously disordered metals. To rule out this scenario, we may first assume our films be homogeneously disordered and calculate the thermal diffusion length $L_T = \sqrt{D\hbar / k_B T}$, where $D = \hbar k_F \ell_e/3m^\ast$ is the diffusion constant, $\hbar$ is the reduced Planck constant, and $m^\ast$ is the effective electron mass. (We take $m^\ast$ to be the free electron mass. \cite{Radermacher1993}) We obtain $L_T < t$ in the regime $T_1  < T < T_2$ for films A11, A14 and A18. This implies that these films, if homogeneously disordered, should be 3D with regard to the AA EEI effect, then the conductivity correction must obey a $\sqrt{T}$ temperature dependence.\cite{Altshuler1985} This is clearly not the case.

In fact, our results must be explained by the electron tunneling conduction in the presence the EEI effect in a granular array. In the strong intergranular coupling regime with $g_T > g_c$, the transport properties of granular metals have been theoretically addressed,\cite{Beloborodov2007} where $g_T$ is a dimensionless tunneling conductance between neighboring grains, $g_c=(1/2 \pi \tilde{d}) \ln (E_c/\tilde{\delta})$ is a critical conductivity, $E_c = e^2/4 \pi \epsilon_0 \epsilon_r a$ is the charging energy with $\epsilon_0$ ($\epsilon_r$) being the permittivity of vacuum (dielectric constant of $I$), and $\tilde{\delta}$ is the mean energy level spacing in a grain. The granular metal theory predicts that, in the temperature regime $T^\ast = g_T \tilde{\delta}/k_B \le T \ll E_c$ ($k_B$ is the Boltzmann constant), incoherent tunneling processes on scales approximately equal to the grain size $a$ play a pivotal role in governing the system conductivity, causing a conductivity correction ($\delta\sigma_1$). As the temperature further decreases to $T < T^\ast$, the coherent electron motion on scales larger than $a$ becomes important, causing another conductivity correction ($\delta\sigma_2$). Thus, the total conductivity is given by $\sigma(T) = \sigma_0 + \delta\sigma_1 (T) + \delta\sigma_2 (T)$, where $\sigma_0 = (2 e^2/\hbar) g_T a^{2-\tilde{d}}$ is the conductivity without the EEI effect,  
\begin{equation}
    \delta \sigma_1 (T) = - \frac{\sigma_0}{2 \pi g_T \tilde{d}} \ln \left [ \frac{g_T E_c}{{\rm max}(k_B T, g_T \tilde{\delta})} \right ],
    \label{dsig_1}
\end{equation}
and, for $\tilde{d}$ = 2,
\begin{equation}
    \delta \sigma_2 (T) = 
    - \frac{\sigma_0}{4 \pi^2 g_T} \ln \left( {\frac{g_T \tilde{\delta}}{k_B T} }\right). 
    \label{dsig_2}
\end{equation}
We note that the $\delta \sigma_2$ term reproduces the AA EEI effect by taking the screening factor $\tilde{F}$ to be zero.

Figures \ref{fig_4} shows that our results can be well described by the predictions (red straight lines) of Eq. (\ref{dsig_1}) in the intermediate $T$ regime. For a quantitative analysis, we may take $E_c \approx 10 k_B T_2$ (Ref. \onlinecite{Sun2010}) and $\epsilon_r \approx 2.5$ to estimate $a$.\footnote{We use $\epsilon_r \approx 2.5$ in this estimation, which is the average of the dielectric constants of vacuum and SiO$_2$.} The $a$ value thus obtained (see Table \ref{table_diff}) is comparable with the corresponding $a_{\rm AFM}$ value. From the fitted $\sigma_0$ and $g_T$ values, the $T^\ast$ values can calculated with $\tilde{\delta} \approx 1/\nu_0 a^3$, where $\nu_0$ is the density of states (DOS) at the Fermi energy of the free-electron model. The $g_c$ values can also be calculated. We obtain $T^\ast \approx T_1$ and $g_T > g_c$ for each film, supporting the interpretation in terms of the $\delta \sigma_1$ correction. Moreover, below $T^\ast$, a crossover to the $\delta\sigma_2$ correction is observed in the two most granular films, A11 and A14, as indicated by the fitted green dashed straight lines given by Eq. (\ref{dsig_2}).\footnote{If we rewrite Eq. (\ref{dsig_2}) into the more familiar AA expression $\Delta \sigma_\square (T) = (e^2/2\pi^2 \hbar) (1 - 3\tilde{F} /4) \ln (T/T_0)$ and fit the data at $T<T_1$ to this equation, we obtain $\tilde{F} \approx -0.20$ ($-0.24$) for film A11 (A14).} Thus, the results of Fig. \ref{fig_4} over the wide temperature from 0.25 K to $T_2$ indicate that our films form 2D, rather than 3D, granular arrays. This conclusion is in line with the 2D granular array structure determined from the AFM and TEM images discussed above.

\begin{table}
\caption{\label{table_diff}
Relevant parameters for three films. $E_c$ is the charging energy, $a$ is the grain diameter calculated from $E_c$, $g_T$ is the dimensionless intergranular tunneling conductance, and $T^\ast$ is a crossover temperature defined in the text.
}

\begin{ruledtabular}
\begin{tabular}{ccccc}
Film & $E_{c}/ k_{B}$ (K) & $a$ (nm) & $g_{T}$ & $T^\ast$ (K) \\ 
\hline 
A11 & 600 & 11 & 2.8 & 6.6 \\
A14 & 370 & 18 & 4.4 & 2.5 \\
A18 & 160 & 42 & 27 & 1.3 \\
\end{tabular}
\end{ruledtabular}
\end{table}

The inset of Fig. \ref{fig_4} shows that films A20, 21 and 23 reveal metallic behavior, with $\sigma_\square$ increasing with decreasing $T$. This result further supports the GHE theory prediction that $R_H$ should peak at $x_q > x_c$. Furthermore, we note that the large $R_H$ enhancement in film A21 cannot be ascribed to the suppression of DOS. The granular metal theory has calculated the corrections to the DOS due to the EEI effect in the presence of granularity.\cite{Efetov2003,Beloborodov2003} For $\tilde{d}$ = 2 and at $T \lesssim T^\ast$, the correction to the DOS at the Fermi energy ($\delta \nu_2$) is given by $\delta \nu_2/\nu_0 = -(16 \pi^2 g_T)^{-1} \left[ \ln (g_T \delta/k_B T) \ln (g_T E_{c}^{4} / k_B T \delta^{3} + 2 \ln ^2 (E_c/\delta)) \right]$. This result indicates that a smaller grain size (a larger $E_c$) and a smaller $g_T$ will cause a larger $\left| \delta \nu_2 \right|$. Consider, for example, film A18. From the values $E_c/k_B \sim$ 160 K and $g_T \sim$ 27, we estimate $\delta \nu_2/\nu_0 \approx -3 \%$, which would lead to a $\sim$\,3\% increase in $R_H$ through the relation $\delta n^\ast/n^\ast \propto \delta \nu_2/\nu_0$, where $\delta n^\ast$ is a correction to $n^\ast$. For the more continuous and less granular film A21, $\delta \nu_2/\nu_0$ should be smaller than that in film A18. This small correction of DOS certainly cannot explain the observed GHE.  

\section{Conclusion}

We have fabricated a series of two-dimensional granular CoSi$_2$ films which encompasses the metal-insulator transition. We observe an enhanced Hall coefficient by a factor of $\approx$\,100, which occurs in a film falling slightly above the metal-insulator transition. We explain this result in terms of the giant Hall effect due to the local quantum-interference effect induced reduction of charge carriers. The two dimensionality of our granular arrays is confirmed by transmission electron microscopy studies and low-temperature transport properties. Granular CoSi$_2$ films are stable under ambient conditions. Their large Hall effect may benefit useful and sensitive applications.

\begin{acknowledgments}
This work was supported by the National Science and Technology Council of Taiwan through grant numbers 110-2112-M-A49-015 and 111-2119-M-007-005 (J.J.L.), and 110-2112-M-A49-033-MY3 (S.S.Y.). S.S.Y. was partly supported by the Center for Emergent Functional Matter Science of NYCU from The Featured Areas Research Center Program within the framework of the Higher Education Sprout Project by the Ministry of Education of Taiwan. The authors acknowledge the use of HRTEM at the Instrument Center of National Tsing Hua University (Taiwan).
\end{acknowledgments}



%

\end{document}